\documentclass[preprint,11pt, authoryear]{elsarticle}
\usepackage{amsfonts, amssymb, amsthm, amsmath, graphicx, caption}
\usepackage[flushleft]{threeparttable} 
\usepackage[colorlinks=true, pdfstartview=FitV, linkcolor=blue, citecolor=red, urlcolor=blue]{hyperref}

\journal{.}

\begin{document}

\begin{frontmatter}


\title{An empirical analysis of the relationships between crude oil, gold
and stock markets 
}
\author{Semei Coronado\corref{cor1}\fnref{label2}}
\ead{semeic@cucea.udg.mx}
\author{Rebeca Jim\'{e}nez-Rodr\'{\i}guez\corref{cor2}\fnref{label3}}
\ead{rebeca.jimenez@usal.es}
\author{Omar Rojas\corref{cor3}\fnref{label4}}
\ead{orojas@up.edu.mx}


\address[label2]{Department of Quantitative Methods, Universidad de Guadalajara, Guadalajara, Mexico}
\address[label3]{Department of Economics, IME, University of Salamanca, Salamanca, Spain}
\address[label4]{School of Business and Economics, Universidad Panamericana, Guadalajara, Mexico}

\begin{abstract}
This paper analyzes the direction of the causality between crude oil, gold and stock markets for the largest economy in the world with respect to such markets, the US. To do so, we apply non-linear Granger causality tests. We find a nonlinear causal relationship among the three markets considered, with the causality going in all directions, when the full sample and different subsamples are considered. However, we find a unidirectional nonlinear causal relationship between the crude oil and gold market (with the causality only going from oil price changes to gold price changes) when the subsample runs from the first date of any year between the mid-1990s and 2001 to last available data (February 5, 2015). The latter result may explain the lack of consensus existing in the literature about the direction of the causal link between the crude oil and gold markets. 
\end{abstract}


\end{frontmatter}

\section{Introduction}
\label{sec:intro}

The crude oil and gold markets are the main representative of the large commodity markets and seem to drive the price of other commodities \citep[see][]{sari2010dynamics}. On the one hand, gold is the leader in the precious metal markets and is considered as an investment asset. Gold is a safe haven to avoid an increase in financial risk \citep[see][]{aggarwal2007psychological}, a store of value \citep[see][]{baur2010gold} and a hedge against inflation \citep[see][]{jaffe1989gold}; consequently it is used as a fundamental investment strategy \citep[see][]{baurMcDermott2010gold}. On the other hand, crude oil is the main source of energy and is also used as an investment asset. Therefore, investors often include one of the two commodities --gold and crude oil-- or both in their investment portfolios as a diversification strategy \citep[see][]{soytas2009world}.

There seems to be a close relationship between the price movements of the two commodity markets, but there is no consensus on the direction of the influence. \citet{baffes2007oil}, \citet{zhang2010crude} and \citet{sari2010dynamics} found that gold prices respond significantly to changes in oil prices. However, there are some authors such as \citet{narayan2010gold} and \citet{wang2013dynamic} that argue that oil and gold prices affect each other.\footnote{ \citet{bampinas2015relationship} found a unidirectional causality from oil prices to gold prices before the 2007/2008 crisis and a
biderectional causality after the crisis.} 
\citet{reboredo2013gold} pointed out the four mechanisms through which crude oil and gold (seen as an investment asset) are linked: a) the increase in oil prices leads to inflationary pressures \citep[see, e.g.][]{hooker2002oil,chen2009revisiting, alvarez2011impact} that induces gold prices to increase since gold is seen as a hedge against inflation; b)\ high oil prices have a negative impact on economic growth \citep[see][]{hamilton2003oil, jimenez2005oil, kilian2008exogenous, cavalcanti2013macroeconomic} and asset values \citep[see][]{reboredo2010nonlinear}, which gives rise to an increase in gold price since it is seen as an alternative asset to store value;\ c) higher oil prices have a positive effect on revenues in net oil exporting countries, which increases their investment in gold to maintain its share in the diversified portfolios and, consequently, gold price increases due to higher gold demand \citep[see][]{melvin1990south};\ and d) when the US dollar depreciates oil prices rise
 \citep[see][]{reboredo2012modelling} and investors may use gold as a safe haven.

Given that oil and gold are used as investment asset,\footnote{%
As was pointed out, gold and oil are often used as a safe haven against the more traditional asset classes such as equities and bonds.} 
they are closely related to the evolution of stock market indices since any influence on decisions about investment portfolios affects the stock market returns \citep[see][]{ciner2013hedges}.

The relationship between changes in oil prices and stock market indices has been widely studied. Authors such as \citet{jones1996oil}, \citet{sadorsky1999oil}, \citet{ciner2001energy}, \citet{park2008oil}, \citet{kilian2009impact}, \citet{ciner2013oil} and \citet{jimenez2015oil} found that an oil price increase has a negative impact on stock returns in oil importing countries,\footnote{
It is worth noting that there are some authors that did not find any significant impact of oil price changes on stock markets \citep[see][]{huang1996energy, apergis2009structural}. } while \citet{bjornland2009oil} and \citet{wang2013oil} found a positive impact of oil price increases on the stock market in oil exporting countries. However, there are fewer authors who have analyzed how gold prices affect stock market indices, and vice versa. \citet{smith2001price}%
\footnote{%
Authors such as \citet{sherman1982gold}, \citet{herbst1983gold} and \citet{jaffe1989gold} had previously studied the role of gold in investment portfolio.}$%
^{,}$\footnote{%
There are authors who state the relative benefits of including gold in the investment portfolios \citep[see][]{sherman1982gold, hillier2006precious, baur2010gold}. \citet{sherman1982gold} indicated that gold has less volatility than stocks and bonds and improves overall portfolio performance.\textrm{\ Hillier \textit{et al}. (2006)} found that portfolios that include precious metals outperform those with standard equity portfolios. \citet{baur2010gold} showed that gold can be considered as a hedge against stocks on average and a safe haven in extreme stock market conditions.} 
studied the relationship between gold prices and the US stock price indices over the 1991-2001 period. He considered four gold prices (three set in London: 10.30 a.m. fixing, 3 p.m. fixing, and closing time; and one set in New York: Handy \& Harmon) and six stock price indices (the Dow Jones Industrial Average, NASDAQ, the New York Stock Exchange, Standard and Poor's 500, Russell 3000, and\ Wilshire 5000) and showed evidence of a short-run relationship, but not a long-run link. He also found evidence of feedback between gold price set in the afternoon fixing and US\ stock price indices by using linear Granger causality test, but unidirectional causality from US\ stock price indices to gold price set in the morning fixing and closing time. 
\citet{bhunia2012association}  analyzed the causal link between gold prices and Indian stock market returns, showing the bidirectional Granger-causality.

The study of the link between the two commodity markets (crude oil and gold) and the stock market indices is of interest to policymakers since the movements in the stock market has an important influence on macroeconomic variables development. To the best of our knowledge, there is no study in the related literature that analyzes the relationships between crude oil, gold and stock markets for the largest economy in the world, the US.\footnote{%
See the last Gross Domestic Product ranking table based on Purchasing Power
Parity provided by the World Bank for 2013
(http://data.worldbank.org/data-catalog/GDP-ranking-table).} 

The contribution of this paper is to extend the literature on the
relationships between the crude oil and gold markets and the Standards and Poor's 500 index by analyzing the direction of the causality and by considering data from the Great Moderation onwards. To do so, we apply the nonlinear Granger causality test for the full sample and for different subsamples in order to analyze the sensitivity of the results to the use of different sample periods. Additionally, we perform the nonlinear Granger causality test for windows of one natural year from 1986 up to 2014 to investigate the causal link within each specific year.

The paper is structured as follows. Section \ref{sec:data} describes the data and the methodology. Section \ref{sec:results} presents the  results and discussion. Section \ref{sec:conclusions} concludes.

\section{Data and methodology}
\label{sec:data}

The empirical sample used for the present study consists of Standard and Poor's 500 daily adjusted closing price (SP500), West Texas Intermediate crude oil spot price (WTI) and Gold Bullion LBM US/Troy Ounce (Gold) from January 2nd, 1986 to February 5th, 2015, with a total of 7351 observations. All data are available from {\em Bloomberg}. These dates were chosen in order to capture different economic moments in the relationships between the series. 

\begin{figure}[!h]
   \centering
    \includegraphics[scale=0.8]{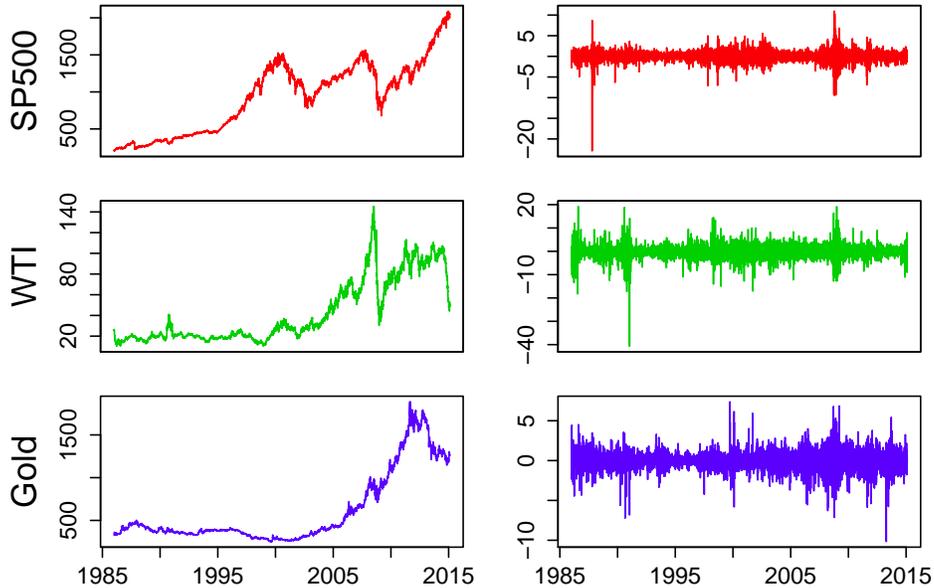}
    \caption[Prices and Returns]{Time series plots of prices, $p_t$, (left) and returns, $r_t$, (right) for SP500 (top), WTI (middle) and Gold (bottom).}
    \label{fig:prices_returns}
\end{figure}

Prices were transformed into series of continuously compounded percentage returns by taking the first differences of the natural logarithm of the prices, i.e. $r_t = 100( \ln(p_t) -\ln(p_{t-1}))$, where $p_t$ is the price on day $t$. We denote the return time series for SP500, WTI and Gold by SP500R, WTIR and GoldR, respectively. Figure \ref{fig:prices_returns} shows the behavior of the prices and returns for each series.  

Table \ref{tab:stats} presents the descriptive statistics of the return time series. The statistics are consistent, as expected, with some of the stylized facts of financial and economic time series \citep[see][]{cont2001}. In particular, the kurtosis indicates that return distributions are leptokurtic. Moreover, the \citet{Jarque1987} statistic confirms returns are not normally distributed.

\begin{table}[!h]
\caption{Descriptive statistics of the return time series}
\label{tab:stats}
\centering
\begin{tabular}{lrrr}
  \hline
Statistic & SP500R & WTIR & GoldR \\ 
  \hline \hline
Mean & $0.03$ & $0.01$ & $0.02$ \\ 
  Min & $-22.90$ & $-40.69$ & $-10.16$ \\ 
  Max & $10.96$ & $19.24$ & $7.38$ \\ 
  Sd & $1.17$ & $2.52$ & $1.00$ \\ 
  Skewness & $-1.28$ & $-0.72$ & $-0.39$ \\ 
  Kurtosis & $30.90$ & $18.24$ & $10.52$ \\ 
  Jarque-Bera & $240345.68$ & $71751.11$ & $17491.18$ \\ 
   \hline
\end{tabular}
\end{table}

We first analyze the stationarity of the variables considered\footnote{It is
worth noting that the causality tests are only valid if the variables have the
same order of integration \citep[see, e.g.][]{papapetrou2001oil}} by applying the
Augmented \citet{dickey1981likelihood} test and the Residual Augmented Least
Squares (RALS) test proposed by \cite{Im2014}, which does not require
either knowledge of a specific density function of the error term or knowledge
of functional forms. 

In addition to stationarity, it is standard to test for linearity of the
asset variables. Thus, we apply the BDS test \citep{bds1996} and the \citet{tsay1986} test to our variables. Whereas the Tsay test is a direct test for
non-linearity of a specific time series, the BDS test is an indirect test.

The aim of the \citet{tsay1986} test is to detect quadratic serial dependence in
the data \citep[see][for further details]{tsay1986}. The BDS test was originally
developed to test for the null hypothesis of independent and identical
distribution ({\em iid}) in order to detect non-random chaotic dynamics, but when it
is applied to the residuals from a fitted univariate linear time series model
the test uncovers any remaining dependence and the presence of an omitted
nonlinear structure. Consequently, if the null hypothesis cannot be rejected,
then the fitted univariate linear model cannot be rejected. However, if the
null hypothesis is rejected, the fitted univariate linear model is
misspecified, and in this sense, it can also be treated as a test for
nonlinearity \citep[see][]{Zivot2006}. There are two main advantages of choosing the BDS test: $(1)$ it has been shown to have more power than other linear and nonlinear tests \citep[see][]{brock1991nonlinear, barnett1997single}; and $(2)$ it is nuisance-parameter-free and does not require any adjustment when applied to fitted model residuals \citep[see][]{f1996nuisance}. See \cite{bds1996} for further details.

We analyze the Granger causality relationship among the variables considered.
Notice that for a strictly stationary bivariate process $\{X_{t},Y_{t}\}$ the
process $\{Y_{t}\}$ is Granger caused by $\{X_{t}\}$ if the past and current
values of $\{X_{t}\}$ contain additional information of future values of
$\{Y_{t}\}$ that is not contained in past and current values of $\{Y_{t}\}$
alone.\footnote{See \citet{granger2001}, \citet{diks2006new} and \citet{Wolski2014} for a formal definition of Granger causality.} Recently, there has been an increase of interest in nonparametric versions of the Granger non-causality hypothesis against linear and nonlinear Granger causality \citep[see][]{hiemstra1994testing, bell1996non, su2008nonparametric}. Given that the linear
Granger causality test might fail to uncover nonlinear causal relationships,
we use the \citet{diks2006new} nonlinear Granger causality test
(hereafter, DP test).

The DP test is a nonparametric Granger causality test based on the use of the correlation integral between time series and based on \citet{baek1992general} but without the assumption of the time series being mutually and individually independent and identically distributed. It has has also been shown to be more  display short-term temporal dependence, since it reduces the over-rejection whenever the null hypothesis is true.

We next describe the DP test closely following the description offered by \citet{diks2006new}. It is denoted by $X_{t}^{l_{X}}=(X_{t-l_{X}+1}
,...,X_{t})$ and $Y_{t}^{l_{Y}}=(Y_{t-l_{Y}+1},...,Y_{t})$ the delay vector of
$X_{t}$ and $Y_{t}$, respectively. The null hypothesis tested is the lack of
causality, that is, that past observations of $X_{t}$ do not contain
additional information about $Y_{t+1}$:
\begin{equation}
H_{0}:Y_{t+1}|(X_{t}^{l_{X}};Y_{t}^{l_{Y}})\sim Y_{t+1}|Y_{t}^{l_{Y}}
\end{equation}
Considering a strictly stationary bivariate time series $\{X_{t},Y_{t}\}$, the
null hypothesis is a statement about the invariant distribution of the
$(l_{X}+l_{Y}+1)-$dimensional vector $W_{t}=(X_{t}^{l_{X}},Y_{t}^{l_{Y}}
,Z_{t})$, with $Z_{t}=Y_{t+1}$. Assuming that the null hypothesis is a
statement about the invariant distribution of $W_{t}$, the time subscript can
be dropped and it can be just written as $W=(X,Y,Z)$. To simplify the test
description, it is assumed that $l_{X}=l_{Y}=1$. Thus, the conditional
distribution of $Z$ given $(X,Y)=(x,y)$ is the same, under the null
hypothesis, as that of $Z$ given $Y=y$. In terms of joint probability density
function, $f_{X,Y,Z}(x,y,z)$, and its marginals, the null hypothesis has to
ensure:
\begin{equation}
\dfrac{f_{X,Y,Z}(x,y,z)}{f_{Y}(y)}=\dfrac{f_{X,Y}(x,y)}{f_{Y}(y)} \cdot \dfrac{f_{Y,Z}(y,z)}{f_{Y}(y)}
\end{equation}
for each vector $(x,y,z)$ in the support of $W$. Thus, it can be stated that
$X$ and $Z$ are independent conditionally on $Y=y$ for each value of $y$ \citep[see][]{bekiros2008relationship, Wolski2014}. \citet{diks2006new} show that the
reformulated null hypothesis implies the $q$ statistic to be noted as
\begin{equation}
q\equiv E\left[  f_{X,Y,Z}(X,Y,Z)f_{Y}(Y)-f_{X,Y}(X,Y)f_{Y,Z}(Y,Z)\right],
\end{equation}
where the proposed estimator for $q$ is:
\begin{equation}
T_{n}(\epsilon_{n})=\frac{(2\epsilon_{n})^{-d_{X}-2d_{Y}-d_{Z}}}
{n(n-1)(n-2)}\sum_{i}\left[  \sum_{k,k\neq i}\sum_{j,j\neq i}\left(
I_{ik}^{XYZ}I_{ij}^{Y}-I_{ik}^{XY}I_{ij}^{YZ}\right)  \right],
\end{equation}
where $I_{ij}^{U}=I(\left\Vert U_{i}-U_{j}\right\Vert <\epsilon_{n})$, with
$I($\textperiodcentered$)$ being an indicator function, $\left\Vert
\text{\textperiodcentered}\right\Vert $ being the maximum norm and
$\epsilon_{n}$ being the bandwidth which depends on the sample size. Denoting
$\hat{f}_{U}(U_{i})$ as the local density estimator of the vector $U$ at
$U_{i}$ ,i.e.,
\begin{equation}
\hat{f}_{U}(U_{i})=\left(  2\epsilon_{n}\right)  ^{-d_{U}}(n-1)^{-1}%
\sum_{j,j\neq i}I_{ij}^{U},
\end{equation}
the test statistic simplifies to
\begin{equation}
T_{n}(\epsilon_{n})=\frac{n-1}{n(n-2)}\sum_{i}\left(  \hat{f}_{X,Y,Z}%
(X_{i},Y_{i},Z_{i})\hat{f}_{Y}(Y_{i})-\hat{f}_{X,Y}(X_{i},Y_{i})\hat{f}%
_{Y,Z}(Y_{i},Z_{i})\right).
\end{equation}
Considering one lag (which implies that $d_{X}=d_{Y}=d_{Z}=1$), for a sequence
with bandwidth $\epsilon_{n}=Cn^{-\beta}$, where $C>0$ and $1/4<\beta<1/3$,
the test statistic satisfies
\begin{equation}
\sqrt{n}\left(  \frac{T_{n}(\epsilon_{n})-q}{S_{n}}\right)  \overset
{D}{\longrightarrow}N(0,1)
\end{equation}
where $S_{n}$ is an estimator of the asymptotic standard error of $T_{n}%
($\textperiodcentered$)$ and $\overset{D}{\longrightarrow}$ denotes
convergence in distribution. We implement a one-tailed version of the test,
rejecting the null hypothesis if the left hand side of the equation $(7)$ is
too large. See \citet{diks2006new} for further details.

\section{Empirical Results}
\label{sec:results}

\begin{table}[!ht]
\centering
 \begin{threeparttable}
\caption{Unit-root tests results}
\label{tab:stationarity}
\begin{tabular}{l r r r}
  \hline 
Series & SP500 & WTI & Gold \\ 
  \hline \hline
Series in levels &  &  &  \\ 
With trend and intercept &  &  &  \\ 
ADF & $-1.51$ & $-2.59$ & $-1.43$ \\ 
RALS & $-1.71$ & $-0.46$ & $-1.06$ \\ 
With drift &  &  &  \\ 
ADF & $0.07$ & $-1.58$ & $-0.22$ \\ 
RALS & $1.32$ & $1.74$ & $-0.15$ \\ 
  \hline
Series in first differences & SP500R & WTIR & GoldR \\ 
With trend and intercept &  &  &  \\ 
 ADF & $-21.36^{***}$ & $-33.77^{***}$ & $-25.85^{***}$ \\ 
 RALS & $-22.66^{***}$ & $-33.50^{***}$ & $-29.09^{***}$ \\ 
With drift &  &  &  \\ 
 ADF & $-21.34^{***}$ & $-33.77^{***}$ & $-25.85^{***} $ \\ 
 RALS & $-22.67^{***}$ & $-33.50^{***}$ & $-29.02^{***}$ \\ 
   \hline
\end{tabular}
\begin{tablenotes}
      \small
      \item Notes: The null hypothesis is that there is  a unit root.  The lag was selected by using Bayesian Information Criteria. One/two/three asterisks mean a $p$-value less than 10\%, 5\% and 1\%, respectively.
    \end{tablenotes}
\end{threeparttable}
\end{table}

Table \ref{tab:stationarity} shows the results of the ADF and the RALS. Both tests fail to reject the null hypothesis that the series in levels are non-stationary and reject
the null hypothesis that the series in first differences are non-stationary at
the 5\% critical level. Thus, the series in levels are integrated of order
one, $I(1)$.

\begin{table}[h!]
\centering
 \begin{threeparttable}
\caption{BDS nonlinearity test results}
\label{tab:BDS}
\begin{tabular}{llrrrr}
  \hline 
Series & $m / \epsilon$ & $0.5\sigma$ & $\sigma$ & $1.5\sigma$ & $2\sigma$ \\ 
  \hline \hline
  & 2 & $12.44^{***}$ & $14.24^{***}$ & $16.58^{***}$ & $19.46^{***}$ \\ 
 SP500R & 3 & $19.14^{***}$ & $21.24^{***}$ & $23.27^{***}$ & $25.50^{***}$ \\ 
  & 4 & $24.42^{***}$ & $25.97^{***}$ & $27.18^{***}$ & $28.73^{***}$ \\ 
  & 2 & $14.57^{***}$ & $17.10^{***}$ & $18.96^{***}$ & $19.32^{***}$ \\ 
 WTIR & 3 & $19.90^{***}$ & $22.29^{***}$ & $24.47^{***}$ & $24.99^{***}$ \\ 
  & 4 & $25.08^{***}$ & $26.20^{***}$ & $27.72^{***}$ & $28.01^{***}$ \\ 
  & 2 & $11.21^{***}$ & $10.24^{***}$ & $10.55^{***}$ & $11.34^{***}$ \\ 
 GoldR & 3 & $16.79^{***}$ & $14.43^{***}$ & $14.21^{***}$ & $14.51^{***}$ \\ 
  & 4 & $22.45^{***}$ & $18.18^{***}$ & $17.39^{***}$ & $17.14^{***}$ \\ 
   \hline
\end{tabular}
\begin{tablenotes}
      \small
      \item Notes: The null hypothesis is that the residuals are iid. One/two/three asterisks mean a $p$-value less than 10\%, 5\% and 1\%, respectively.
    \end{tablenotes}
\end{threeparttable}
\end{table}

We perform the BDS test on the residuals of the return time series for
dimensions up to 4 and values of $\epsilon$ equal to 0.5$\sigma_{x}$,
$\sigma_{x}$, 1.5$\sigma_{x}$, where $\sigma_{x}$ represents the standard
deviation of the return time series $x_{t}$. The results of the tests show
the rejection of the null hypothesis in all cases (see Table \ref{tab:BDS}), from which the nonlinearity of the series can be inferred.

\begin{table}[!ht]
\centering
 \begin{threeparttable}
\caption{Tsay nonlinearity test results}
\label{tab:tsay}
\begin{tabular}{c l l l}
  \hline
 Lag & SP500R & WTIR & GoldR \\ 
  \hline \hline
   1 & $23.81^{***}$ & $16.63^{***}$ & $1.11$ \\ 
   2 & $12.03^{***}$ & $12.12^{***}$ & $1.74$ \\ 
   3 & $10.16^{***}$ & $13.66^{***}$ & $2.60^{**}$ \\ 
   4 & $11.08^{***}$ & $10.29^{***}$ & $2.21^{**}$ \\ 
   5 & $10.11^{***}$ & $8.13^{***}$ & $1.91^{**}$ \\ 
   6 & $8.49^{***}$ & $7.36^{***}$ & $2.00^{***}$ \\ 
   7 & $7.49^{***}$ & $7.29^{***}$ & $2.27^{***}$ \\ 
   \hline
\end{tabular}
\begin{tablenotes}
      \small
      \item Note: The null is that the series are linear. One/two/three asterisks mean a $p$-value less than 10\%, 5\% and 1\%, respectively.
    \end{tablenotes}
\end{threeparttable}
\end{table}

Table \ref{tab:tsay} shows the results of the \citet{tsay1986} test, showing that we reject the
null hypothesis of linearity and confirming results found with the BDS test.
Therefore, there seems to be a nonlinear univariate structure behind all our
time series. This information is considered and we use
the nonlinear Granger-causality test proposed by \citet{diks2006new}.

We apply the DP test to the delinearized series. The series have been delinearized by using a VAR filter, whose lag length is chosen on the basis of Bayesian Information Criterion. As \citet{hiemstra1994testing} and \citet{bampinas2015relationship} state "\textit{by removing linear predictive power with a VAR model, any causal linkage from one residual series of the VAR model to another can be considered as nonlinear predictive power}". We perform the DP test for the full sample and for different subsamples to investigate the sensitivity of such results to the use of different sample periods. Additionally, we apply the DP test for windows of one natural year from 1986 up to 2014 in order to analyse the causal link within each specific year.

Table \ref{tab:DP-full} presents the results of DP\ test for the full sample (January 2, 1986 - February 5, 2015) in a compact way following the simplifying notation of \cite{bekiros2008relationship}, who consider one/two/three asterisks to indicate that the corresponding p-value of a test is lower than 10\%, 5\% and 1\%, respectively. Directional causalities will be denoted by the functional representation $\nrightarrow$, {\em i.e.}, $x\nrightarrow y$ means that the VAR filtered $r_{x}$ series does not Granger cause the VAR filtered $r_{y}$ series. ($ \cdot,\cdot,\cdot$) denotes the value of $p$ for the $VAR(p)$ filter of the series \{$x,y$\},
\{$x,z$\} and \{$y,z$\}. Finally, $x$ represents SP500R, $y$ refers to WTIR and $z$ is GoldR.

\begin{table}[!ht]
\centering
 \begin{threeparttable}
\caption{Nonlinear Granger causality test results for the full sample}
\label{tab:DP-full} 
\begin{tabular}{r |  ll  |  ll  |  ll}
  \hline 
Lag & $x \nrightarrow y$ & $y \nrightarrow x$ & $x \nrightarrow z$ & $z \nrightarrow x$ & $y \nrightarrow z$ & $z \nrightarrow y$ \\ 
  \hline \hline
   1 & $***$ & $***$ & $***$ & $***$ & $***$ & $***$ \\ 
   2 & $***$ & $***$ & $***$ & $***$ & $***$ & $***$ \\ 
   3 & $***$ & $***$ & $***$ & $***$ & $**$ & $**$ \\ 
   4 & $***$ & $***$ & $***$ & $**$ & $*$ & $**$ \\ 
   5 & $***$ & $***$ & $***$ & $*$ & $*$ & $***$ \\ 
   \hline
\end{tabular}
\begin{tablenotes}
      \small
      \item Note: Note: One/two/three asterisks mean a $p$-value less than 10\%, 5\% and 1\%, respectively. The null hypothesis is that $r_x$ does not Granger cause $r_y$ (i.e.,  $x \nrightarrow y$). Denoting SP500R, WTIR and GoldR as $x$, $y$ and $z$, respectively.
    \end{tablenotes}
\end{threeparttable}
\end{table}

Table \ref{tab:DP-full} indicates that the three markets considered (the crude oil, gold and stock markets) are interrelated, with the causality going in all directions. The bidirectional causality between the price movements of the two commodity markets is in concordance with \cite{narayan2010gold} and \cite{wang2013dynamic}, showing the mutual influence of the two investment assets since the Great Moderation. Moreover, the bidirectional causality between the Standard and Poor's 500 returns and changes in the price of the two commodities implies that movements in S\&P's 500 index may be monitored by observing changes in commodity prices and vice versa. The feedback relationship between the crude oil and stock markets is in line with \cite{ciner2001energy}, among others. The bidirectional causal relationship between the gold market and the US stock price index is very relevant and has been only found by \cite{smith2001price} for a specific gold price (3 p.m. fixing).

\begin{table}[!t]
\centering
 \begin{threeparttable}
\caption{Nonlinear Granger causality test results for rolling windows from 1986 to the ending year}
\label{tab:DP-up} 
\begin{tabular}{c | cc  | ll  |  ll  |  ll}
  \hline
Ending year & $\epsilon$ & sample size & $x \nrightarrow y$ & $y \nrightarrow x$ & $x \nrightarrow z$ & $z \nrightarrow x$ & $y \nrightarrow z$ & $z \nrightarrow y$ \\ 
  \hline \hline
  1991 & 1.15 & 1517 & * &  &  &  & * & * \\ 
  1992 & 1.05 & 1773 & ** & * & * & * & ** & * \\ 
  1993 & 1.00 & 2032 & ** & * & * & * & ** & ** \\ 
  1994 & 0.97 & 2289 & *** & * & * & ** & ** & ** \\ 
  1995 & 0.96 & 2541 & *** & ** & * & * & *** & *** \\ 
  1996 & 0.94 & 2795 & *** & ** & ** & * & * & * \\ 
  1997 & 0.92 & 3048 & *** & ** & ** & ** & * & * \\ 
  1998 & 0.90 & 3300 & ** & * & * & * & * & ** \\ 
  1999 & 0.88 & 3552 & *** & * & * & ** & ** & ** \\ 
  2000 & 0.86 & 3804 & *** & * & ** & ** & * & ** \\ 
  2001 & 0.84 & 4053 & *** & * & ** & *** & * & * \\ 
  2002 & 0.82 & 4305 & *** & * & *** & ** & * & ** \\ 
  2003 & 0.80 & 4557 & *** & * & ** & *** & ** & *** \\ 
  2004 & 0.78 & 4809 & *** & * & ** & ** & ** & *** \\ 
  2005 & 0.76 & 5061 & *** & ** & ** & ** & ** & *** \\ 
  2006 & 0.75 & 5312 & *** & * & * & ** & ** & *** \\ 
  2007 & 0.74 & 5563 & *** & * & * & ** & ** & ** \\ 
  2008 & 0.74 & 5816 & *** & ** & ** & ** & ** & *** \\ 
  2009 & 0.73 & 6068 & *** & ** & *** & *** & ** & *** \\ 
  2010 & 0.73 & 6320 & *** & * & *** & *** & * & * \\ 
  2011 & 0.72 & 6572 & *** & ** & ** & ** & ** & *** \\ 
  2012 & 0.71 & 6822 & *** & * & *** & *** & * & *** \\ 
  2013 & 0.70 & 7074 & *** & ** & *** & *** & * & *** \\ 
  2014 & 0.69 & 7326 & *** & ** & *** & *** & * & * \\ 
   \hline
\end{tabular}
\begin{tablenotes}
      \small
      \item Note: One, two and three asterisks mean a $p$-value less than 10\%, 5\% and 1\%, respectively. The null hypothesis is that $r_x$ does not Granger cause $r_y$ (i.e., $x \nrightarrow y$). Denoting SP500R, WTIR and GoldR as $x$, $y$ and $z$, respectively. Even though the DP test is run considering different lags ($l = 1, 2, 3, 4, 5$), it is only reported for each sample the lowest significance level (*,** or ***) for the cases in which causality appears at every lag considered. If no asterisk appears for a particular sample, it means that the causality may exist at some specific lag but not at all possible lags considered.
    \end{tablenotes}
\end{threeparttable}
\end{table}

To analyze whether the results obtained depends on the sample period considered we apply the DP test to different subsamples, with subsamples containing a minimum of five years of observations. Whereas Table \ref{tab:DP-up} presents the results for the subsamples that run from January 2, 1986 to the last available observation of the year indicated under the notation "ending year", Table \ref{tab:DP-down} shows the results for the subsamples that run from the first available observation of the year indicated under the notation "starting year" to February 5, 2015. Both Tables only report for each sample the lowest significance level (*,** or ***) for the cases in which causality appears at every lag considered. If no asterisk appears for a particular sample, it means that the causality may exist at some specific lag but not at all possible lags considered.

\begin{table}[!ht]
\centering
 \begin{threeparttable}
\caption{Nonlinear Granger causality test results for rolling windows from the starting year to 2014}
\label{tab:DP-down} 
\begin{tabular}{c | cc  | ll  |  ll  |  ll}
  \hline
Starting year & $\epsilon$ & sample size &  $x \nrightarrow y$ & $y \nrightarrow x$ & $x \nrightarrow z$ & $z \nrightarrow x$ & $y \nrightarrow z$ & $z \nrightarrow y$ \\ 
  \hline \hline
1987 & 0.70 & 7099 & ** & * & *** & *** & ** & * \\ 
  1988 & 0.71 & 6846 & *** & ** & *** & *** & * & *** \\ 
  1989 & 0.72 & 6592 & *** & ** & *** & *** & * & ** \\ 
  1990 & 0.72 & 6340 & *** & ** & *** & *** & * & ** \\ 
  1991 & 0.73 & 6086 & *** & * & *** & *** & * & * \\ 
  1992 & 0.74 & 5833 & *** & * & *** & *** & ** & * \\ 
  1993 & 0.74 & 5577 & *** & ** & *** & *** & *** & * \\ 
  1994 & 0.75 & 5318 & ** & * & *** & *** & * & * \\ 
  1995 & 0.76 & 5061 & ** & * & *** & *** & *** & * \\ 
  1996 & 0.77 & 4809 & ** & ** & *** & ** & * &  \\ 
  1997 & 0.79 & 4555 & *** &  & *** & *** & ** &  \\ 
  1998 & 0.82 & 4302 & *** & ** & *** & *** &  &  \\ 
  1999 & 0.84 & 4050 & *** & ** & *** & *** & ** &  \\ 
  2000 & 0.86 & 3798 & ** & ** & *** & *** & ** &  \\ 
  2001 & 0.88 & 3546 & *** & * & *** & *** & ** &  \\ 
  2002 & 0.90 & 3297 & ** & ** & *** & ** & *** & * \\ 
  2003 & 0.85 & 3045 & ** & ** & ** & ** & * & * \\ 
  2004 & 0.97 & 2793 & * & * & *** & ** & * & * \\ 
  2005 & 1.00 & 2541 & ** & ** & *** & ** & ** & * \\ 
  2006 & 0.98 & 2289 & ** & * & *** & *** & * & ** \\ 
  2007 & 1.00 & 2038 & *** & ** & *** & *** & * & * \\ 
  2008 & 1.05 & 1787 & * & ** & *** & *** & * & * \\ 
  2009 & 1.09 & 1534 & * & * & *** & *** & * & * \\ 
   \hline
\end{tabular}
\begin{tablenotes}
      \small
      \item Note: One, two and three asterisks mean a $p$-value less than 10\%, 5\% and 1\%, respectively. The null hypothesis is that $r_x$ does not Granger cause $r_y$ (i.e., $x \nrightarrow y$). Denoting SP500R, WTIR and GoldR as $x$, $y$ and $z$, respectively. Even though the DP test is run considering different lags ($l = 1, 2, 3, 4, 5$), it is only reported for each sample the lowest significance level (*,** or ***) for the cases in which causality appears at every lag considered. If no asterisk appears for a particular sample, it means that the causality may exist at some specific lag but not at all possible lags considered.
    \end{tablenotes}
\end{threeparttable}
\end{table}

Tables \ref{tab:DP-up} and \ref{tab:DP-down} show that we reject the null hypothesis of that S\&P's 500 returns do not Granger cause oil price changes at the 1\% critical level for most subsamples. We also reject that oil price changes do not Granger cause S\&P's 500 returns at a 5\% critical level for an important number of subsamples. Thus, the bidirectional causality between the crude oil and stock markets found for the full sample is verified when different subsamples are considered, although the causality from oil price changes to S\&P's 500 returns seems to weaken depending on when the sample starts and ends.

\begin{table}[!ht]
\centering
 \begin{threeparttable}
\caption{Nonlinear Granger causality test results by year}
\label{tab:DP-year} 
\begin{tabular}{r | ll  |  ll  |  ll}
  \hline
years & $x \nrightarrow y$ & $y \nrightarrow x$ & $x \nrightarrow z$ & $z \nrightarrow x$ & $y \nrightarrow z$ & $z \nrightarrow y$ \\ 
  \hline \hline
1986 &  &  &  &  &  &  \\ 
  1987 &  &  & * &  &  &  \\ 
  1989 &  &  & * &  &  & * \\ 
  1990 & * & ** & ** &  & * &  \\ 
  1991 &  & ** & * &  & ** & ** \\ 
  1992 &  & * & * &  &  &  \\ 
  1993 &  &  &  &  &  &  \\ 
  1994 &  &  & * &  & * & * \\ 
  1995 &  &  &  &  &  & * \\ 
  1996 &  &  &  & * &  &  \\ 
  1997 &  &  & * &  &  & * \\ 
  1998 &  &  &  &  &  &  \\ 
  1999 &  &  &  &  & * &  \\ 
  2000 & * &  &  &  &  &  \\ 
  2001 &  &  &  &  &  &  \\ 
  2002 &  &  &  &  & * &  \\ 
  2003 & * &  & * &  &  & * \\ 
  2004 &  &  & * &  & * & ** \\ 
  2005 & * &  &  &  & * &  \\ 
  2006 &  &  & * & * &  &  \\ 
  2007 & * &  & * &  &  & * \\ 
  2008 & *** & ** & ** & * & ** & * \\ 
  2009 & ** & * & * & * & * & ** \\ 
  2010 & * & * &  &  &  & * \\ 
  2011 &  & * & ** & * &  &  \\ 
  2012 & * &  &  &  &  &  \\ 
  2013 &  & * & * & * &  & * \\ 
  2014 & ** & ** &  &  & * &  \\ 
   \hline
\end{tabular}
\begin{tablenotes}
      \small
      \item Note: One, two and three asterisks mean a $p$-value less than 10\%, 5\% and 1\%, respectively. The null hypothesis is that $r_x$ does not Granger cause $r_y$ (i.e., $x \nrightarrow y$). Denoting SP500R, WTIR and GoldR as $x$, $y$ and $z$, respectively. Even though the DP test is run considering different lags ($l = 1, 2, 3, 4, 5$), it is only reported for each sample the lowest significance level (*,** or ***) for the cases in which causality appears at every lag considered. If no asterisk appears for a particular sample, it means that the causality may exist at some specific lag but not at all possible lags considered.
    \end{tablenotes}
\end{threeparttable}
\end{table}

Tables \ref{tab:DP-up} and \ref{tab:DP-down} also indicate the rejection of the null hypothesis when the returns of the gold and stock markets are considered, showing so a bidirectional causality for almost all subsamples and confirming the results obtained for the full sample. Additionally, these tables reveal the feedback relationship between the crude oil and gold markets when the sample starts in January 2, 1986 and ends in the last available observation of any year beyon 1991. However, a unilateral Granger causality was found from oil price changes to gold price changes when the sample starts in the first available observation of any year between the mid-1990s and 2001 and ends in February 5, 2015. This may explain why some authors \citep[see, e.g.][]{zhang2010crude} find a unilateral causality.

Finally, Table \ref{tab:DP-year} reports the results of the DP test for windows of one natural year from 1986 up to 2014, although these results have to be considered with caution since the DP test is an asymptotic test. Table \ref{tab:DP-year} shows no evidence of a causality within most of the years considered. The main exception to this is for the years 2008 and 2009 (the period of global financial crisis), where the three markets appear interrelated. These causal relationships might be due to the fact that gold is considered a safe haven for investors in times of financial crisis and economic instability and also due to investors seeking refuge on gold and other commodities such as oil and energy derivatives after the collapse of the real state market.

\section{Conclusions}
\label{sec:conclusions}

This paper provides new evidence on a nonlinear causal link among the three markets considered (with the causality going in all directions) for the full sample, for subsamples starting in January 2, 1986 and ending in the last available observation of any year beyond 1991 and for subsamples starting in the first available observation of any year beyond 1987 and ending in February 5, 2015. The only exception to this is the existence of a nonlinear unidirectional causal relationship between the crude oil and gold market (with the causality only going from oil price changes to gold price changes) for subsamples that going from the first date of any year between the mid-1990s and 2001 to February 5, 2015. Therefore, the causality between the price movements of the crude oil and gold markets seems highly dependent on the sample used, which may explain the contradictory results found in the related literature.

The causal link found among the three markets implies that changes in the S\&P's 500 index may be monitored by observing changes in the returns of the two commodity markets considered (and vice versa), which is valuable for policymakers





\section*{Acknowledgments}
Rebeca Jim\'{e}nez-Rodr\'{\i}guez acknowledges support from the Ministerio de Econom\'{\i}a y Competitividad under Research Grant ECO2012-38860-C02-01. 





\section*{References}


\end{document}